\documentclass[11pt]{article}
\usepackage{amsmath,amsthm,amssymb,amscd}
\usepackage{xurl}
\usepackage{hyperref}
\usepackage{geometry}
\usepackage{tikz-cd}
\usepackage{booktabs} 
\usepackage{graphicx} 
\usepackage{multirow}
\usepackage{tikz}
\usepackage{booktabs}
\usetikzlibrary{calc,positioning,3d}
\usepackage[normalem]{ulem} 
\usepackage{imakeidx}
\makeindex
\usepackage[numbers]{natbib}
\usepackage{float}

\usepackage{subfigure}
\usepackage{caption}
\usepackage{svg}

\usepackage{makecell} 
\usepackage{todonotes}

\newtheorem{theorem}{Theorem}[section]
\newtheorem{lemma}[theorem]{Lemma}
\newtheorem{proposition}[theorem]{Proposition}
\newtheorem{corollary}[theorem]{Corollary}
\theoremstyle{definition}
\newtheorem{definition}[theorem]{Definition}
\newtheorem{example}[theorem]{Example}
\newtheorem{remark}[theorem]{Remark}

\numberwithin{equation}{section}

\title{FPCN: Fragment Path Complex Networks for Viral Genome Classification}
\author{
Alice Wachira$^{1}$, Xiang Liu$^{1}$ and
Guo-Wei Wei$^{1,2,3}$%
\thanks{Corresponding author: Guo-Wei Wei (guowei.wei@uga.edu).}
\\
\small $^{1}$Department of Mathematics, University of Georgia,
Athens, GA 30602, USA.
\\
\small $^{2}$Department of Biochemistry and Molecular Biology,
University of Georgia, Athens, GA 30602, USA.
\\
\small $^{3}$School of Computing,
University of Georgia, Athens, GA 30602, USA.
}
	
	\date{}
\begin{document}
 \maketitle 

\begin{abstract}	    
Accurate classification of viral families is essential to understanding genomic diversity, but rapid evolution and heterogeneous genome structure complicate reliable assignment. Newly discovered viruses, particularly those from underrepresented ecosystems, expand the known diversity of viral genomes. This growing diversity poses challenges for classification methods that rely on similarity to existing references. We present Fragment Path Complex Networks (FPCN), a framework that represents viral genomes as sequences of ordered fragments at various scales. FPCN captures contextual sequence information within each fragment and uses path complexes to model complex relationships between neighboring fragments. We evaluated FPCN on well-established NCBI benchmarks, and FPCN achieved higher classification accuracy than competing methods. Further evaluations on newly released datasets revealed FPCN's ability to generalize to genomes that were not presented in the training. Together, these results position FPCN as an effective tool for classifying viral genomes within known families across benchmark datasets and newly released genomes.

\end{abstract}



\clearpage
 \section{Introduction} \label{sec:introduction}
Viruses exhibit extensive genomic diversity and rapid evolutionary dynamics, presenting fundamental challenges for their accurate classification. Reliable taxonomic assignment of viral genome sequences is crucial for identifying novel viruses, tracking pathogen emergence, and supporting public health and environmental surveillance \cite{he2025vitax,jiang2023virus}. The rapid growth of microbial and viral genomic data has further increased the need for efficient and scalable analytical methods \cite{martin2025protein}. However, rare and highly divergent viruses remain challenging to identify because of substantial genomic differences from known reference families \cite{de2026ensemble}. These challenges motivate computational frameworks capable of learning informative genomic representations while preserving the sequential structure of viral genomes. Such computational approaches are particularly important for characterizing rapidly evolving viruses and emerging variants \cite{chen2022omicron,wang2026variant}.

Classical viral classification methods have relied mostly on sequence alignment and similarity to known reference genomes. Alignment-based methods compare query sequences with known references, often using multiple sequence alignment tools such as MUSCLE and ClustalW or reference-supported voting schemes to infer taxonomic labels \cite{edgar2004muscle,thompson2003multiple,sadad2023classification}. Despite their success, these methods are highly dependent on the quality of reference databases, which limits performance for novel or highly divergent viruses \cite{sadad2023classification,suwayyid2026cakr}. Alignment-based methods are computationally demanding for large datasets and long genomes. Moreover, they may not be applicable to genomes with dramatically different sequences  \cite{zielezinski2017alignment,shaukat2023comparative,azevedo2024deepvirusclassifier}. To overcome these challenges, alignment-free methods enable efficient genome comparison between diverse genome families \cite{wu2019magnus,van2025alignment}. However, these techniques may discard important genomic structure information. For example, $k$-mer based approaches can overlook repetitive organization and other higher-order sequence relationships \cite{charvel2026reskmer}.

Subsequence-based methods can efficiently encode viral genomes using short nucleotide patterns \cite{wang2018virus, yu2026knvresgat}. Recent research has supplemented traditional frequency-based methods with positional, structural, topological, and algebraic information. Dimensionality reduction coupled with clustering has been used to characterize genomic variation in large-scale influenza mutation datasets \cite{walden2025dimensionality}. Hozumi and Wei introduced $k$-mer topology, a multiscale framework that uses persistent homology or persistent Laplacians to describe the topological structure and evolution of $k$-mer sequence space \cite{hozumi2026revealing}. This approach not only accounts for $k$-mer frequency but also encodes relative $k$-mer position information \cite{liu2026topological2,liu2026topological}.    
Suwayyid et al. extended this approach by developing a commutative algebra $k$-mer representation for genetic variant classification, phylogenetic analysis, and viral classification \cite{suwayyid2026cakr}. This approach outperforms other competing methods. 

Large language models typically accept inputs of limited length and therefore cannot be directly applied to long genomic sequences. Other machine learning approaches, such as those based on $k$-mer frequency profiles, provide an alternative by representing genomes through compositional features. However, these methods can be sensitive to sequencing artifacts, compositional bias, and predefined feature choices \cite{sinno2025charting,wang2024transginmer}. Deep learning has improved automated sequence representation learning. Nevertheless, many deep learning algorithms rely on window-based processing or simply pooling operations, which often lose informative long-range and multiscale genomic dependencies \cite{miao2022virtifier}. Topological deep learning (TDL), introduced in 2017 \cite{cang2017topologynet}, combines topological features with neural networks to capture structural information in biomolecular data. Recent TDL approaches incorporate high-order topological relationships into neural networks \cite{li2024path, ebli2020simplicial, papamarkou2024position}.
TDL techniques have been used to predict mutation-induced changes in protein-protein binding affinity, characterize mutations linked to cross-species transmission, and to assess the potential dominance of emerging SARS-CoV-2 variants \cite{wee2025rapid,wee2024preventing,chen2022omicron}. Current TDL approaches still face challenges in integrating contextual sequence information and ordered fragment dependencies on multiple genomic scales for viral classification \cite{su2025topological,wee2025review,papamarkou2024position}.

 Here, we present Fragment Path Complex Networks (FPCN), a framework for viral family classification that integrates contextual sequence information with ordered fragment relationships at multiple scales. Rather than using whole-genome sequence representations, FPCN employs a multiscale fragment technique to partition genomes into equal fragments using four scales. The framework uses DNABERT-2 to encode the sequence context within each fragment. At each scale, a path complex represents individual fragments, adjacent pairs and consecutive triples, explicitly retaining relationships that independent fragment encoding or pooling can overlook. A shared path complex neural network (PCNN) learns from these ordered relationships, and learned fusion integrates complementary representations across scales, reducing reliance on a single fragment resolution. We evaluate FPCN on established NCBI benchmarks and newly collected genomes using cross-validation and temporal evaluations on known viral families. FPCN significantly outperforms all benchmark methods. This establishes FPCN as an efficient tool for accurate assignment of viral families.

\section{Results} \label{sec:results}

\subsection{Model Overview}\label{subsec:model overview}
\begin{figure}[!t]
    \centering
    \includegraphics[width=1\linewidth]{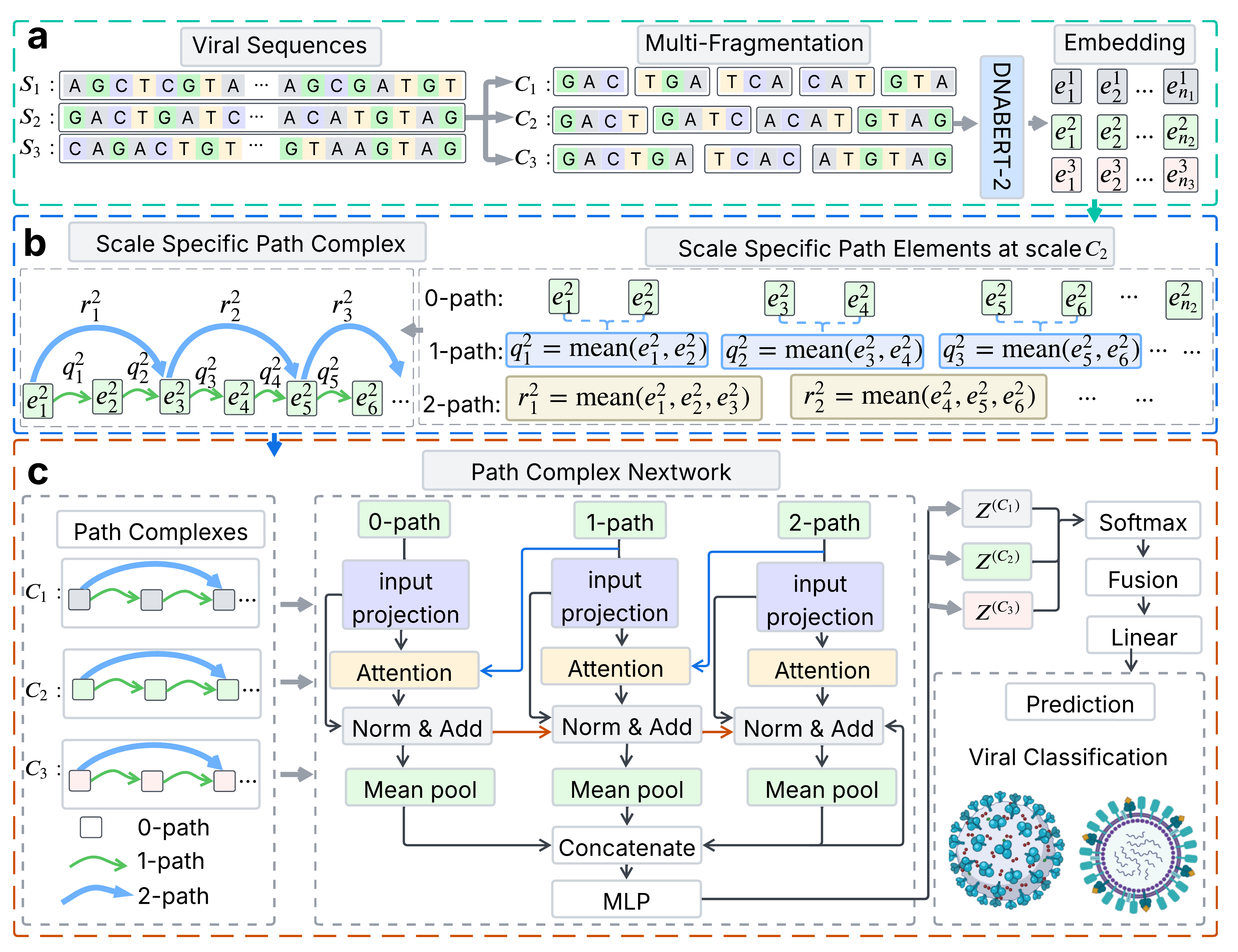}
    \caption{Architecture of the fragment path complex networks (FPCN) for viral family classification.
(a) Each viral genome is partitioned at various fragment scales, and each segment is encoded using pretrained DNABERT-2. (b) For each scale, segment embeddings define 0-paths, adjacent segment pairs and consecutive segment triples define 1-paths and 2-paths, respectively, producing a scale-specific path complex. (c) A shared path complex neural network updates the three path ranks through attention, lifting, and residual normalization. The updated features are pooled to obtain one genome embedding per scale. Learned softmax weights fuse the scale-specific embeddings before performing the viral family classification.}
    \label{fig:flow_chart}
\end{figure}

The classification of viral families requires an approach that preserves both the local sequence patterns and global genome organization. FPCN is designed to capture genome structures by combining multiscale sequence fragmentation, pretrained DNABERT-2 embeddings, and path complex neural learning. As shown in Fig.~\ref{fig:flow_chart}a, each viral genome is partitioned into fragments at different scales to represent sequence organization at different resolutions. The resulting fragments are encoded with DNABERT-2, a pretrained genomic language model that converts nucleotide sequences into contextual embeddings and captures sequence composition and dependency patterns. In Fig.~\ref{fig:flow_chart}b, the fragment embeddings are organized into scale-specific path complexes that preserve ordered relationships among fragments within each genome scale. Neighboring fragments are used to define $0$-paths, while $1$-paths and $2$-paths are defined using adjacent pairs and triples, respectively. These path elements preserve the local order within each scale-specific path complex. As shown in Fig.~\ref{fig:flow_chart}c, the features of each scale-specific path complex are passed to a PCNN for the classification of the viral family. The PCNN updates the $0$-path, $1$-path, and $2$-path features through attention message passing. The updated features are then pooled into one genome embedding per scale. Finally, the embeddings are combined from all scales and passed to a linear classifier to predict the viral family.

\subsection{Datasets}\label{subsec:dataset}
The data considered in this study comprise previous NCBI benchmarks and newly constructed NCBI 2026 datasets. The benchmark datasets include NCBI 2020, NCBI 2022, NCBI 2024, and NCBI 2024 All, which were adopted from previous studies \cite{hozumi2026revealing,suwayyid2026cakr}. A detailed description of the dataset can be found in the Supporting Information(Section 1).  To assess performance on newly released genomes, we collected complete RefSeq viral genomes from NCBI Virus and retrieved their taxonomic annotations from NCBI Taxonomy as of August 10, 2026. Viral family labels were assigned using family-level taxonomic annotations, and families represented by a single genome were excluded to ensure minimal within-class representation. The resulting two datasets were denoted as NCBI 2026 
and  NCBI 2026 All. The former is a filtered dataset containing genomes with valid nucleotides, while the latter contains all viral genomes collected from retained families.

Next, we constructed temporal test sets by comparing accessions between the 2024 and 2026 NCBI datasets. Genomes with accessions already present in the corresponding 2024 datasets were removed from the 2026 datasets. The remaining genomes were treated as newly released 2026 genomes. For the closed-set temporal evaluation, we retained only newly released genomes belonging to viral families represented in the corresponding 2024 datasets. 

Performance was evaluated using accuracy, balanced accuracy (BA), macro-F1 score (F1), recall, and precision. Macro-averaged metrics were used to account for class imbalance by giving equal weight to each viral family. All benchmark results reported for comparison were taken from \cite{suwayyid2026cakr}. To improve the robustness of the evaluation, each experiment was repeated with $30$ different random seeds, and the mean performance was reported as the final result.

\subsection{Performance Comparison with Existing Methods on NCBI Benchmarks}
FPCN was first evaluated against existing methods on established NCBI benchmark datasets. The existing methods include CAKR \cite{suwayyid2026cakr}, natural vector method (NVM) \cite{sun2021geometric}, feature frequency profiles with Jensen-Shannon divergence (FFP-JS) \cite{sims2009alignment} feature frequency profiles with Kullback-Leibler divergence (FFP-KL) \cite{jun2010whole}, Fourier power spectrum (FPS) \cite{hoang2015new} and Markov k-string (MKS) \cite{qi2004whole}. We retained viral families with at least three genome sequences to ensure minimal family-level representation across the $75:15:15$ training, validation, and test splits. 

\begin{figure}[!t]
    \centering
    \includegraphics[width=0.7\linewidth]{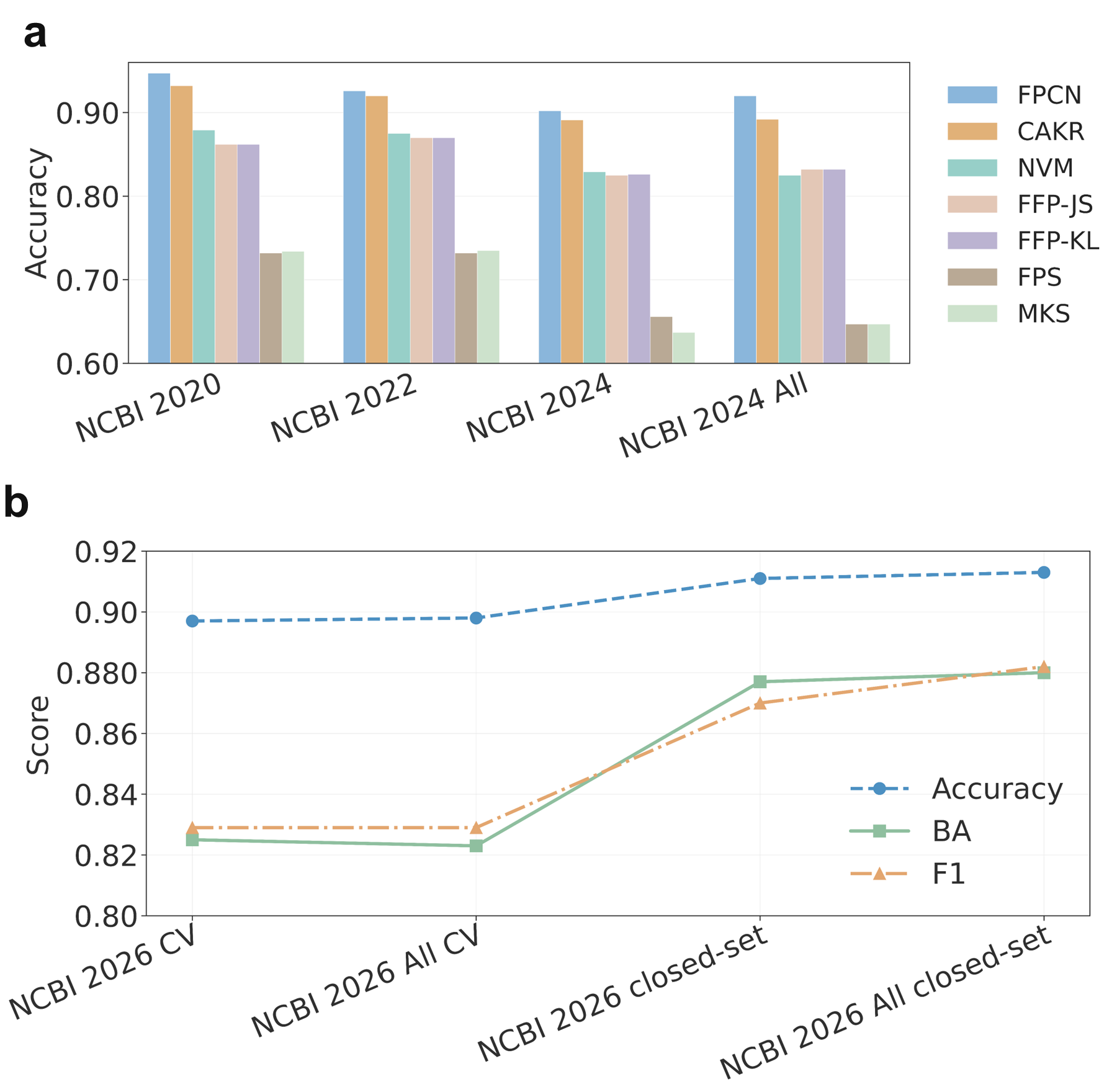}
    \caption{Performance evaluation on NCBI genome datasets. \textbf{a} Accuracy comparison between FPCN and existing methods on established NCBI benchmarks. \textbf{b} FPCN performance on newly collected NCBI 2026 genomes.}
    \label{fig:fpcn_performance}
\end{figure}

The performance of the proposed FPCN model is compared with existing state-of-the-art methods in terms of accuracy in Fig.~\ref{fig:fpcn_performance}a. FPCN achieved the highest classification accuracy on all datasets as compared to other models. In particular, FPCN attained the highest accuracy on NCBI 2020, reaching an accuracy $0.947$ compared to $0.932$ for CAKR, the second-best model. The performance of FPCN exceeded CAKR, by margins ranging from $0.6\%$ on NCBI 2022 to $2.8\%$ on NCBI 2024 All. The greatest improvement was observed on NCBI 2024 All, one of the large benchmarks in this comparison. This suggests that the multiscale path complex representation is particularly useful when the dataset contains a larger and more diverse collection of viral genomes.

\subsection{Performance Evaluation on Newly Collected Viral Genomes}
FPCN was evaluated on newly collected viral genomes in two experiments. First, a five-fold cross-validation was performed separately on NCBI 2026 and NCBI 2026 All datasets. The refined NCBI 2026 dataset contained $14,149$ genomes from $272$ families, and NCBI 2026 All had $15,098$ genomes from $278$ families. For cross-validation evaluation, viral families represented by fewer than ten genomes were excluded to ensure sufficient representation across the training and test folds. 

In the second experiment, a temporal closed-set evaluation was performed to test our model on genomes released after the corresponding NCBI 2024 datasets. FPCN was trained on the filtered NCBI 2024 and NCBI 2024 All datasets and evaluated on NCBI 2026 and NCBI 2026 All closed-set, respectively. The temporal closed-set experiment evaluated whether FPCN could generalize to newly added genomes from known viral families. The NCBI $2026$ All temporal closet-set contained $655$ genomes from $52$ families, while the corresponding NCBI $2026$ closet-set contained $620$ genomes from $50$ families. 

Table~\ref{tab:fpcn_results_2026} summarizes the performance of FPCN on the NCBI 2026 evaluation experiments. Fig.~\ref{fig:fpcn_performance}b shows the corresponding performance trends. For five-fold cross-validation experiment, FPCN achieved similar performance on the refined and complete datasets, with accuracies of $0.897$ on NCBI 2026 CV and $0.898$ on NCBI 2026 All CV with macro-F1 scores of $0.829$ for both. In the temporal closed-set evaluation, the performance increased to an accuracy of $0.911$ on NCBI 2026 closed-set and an accuracy of $0.913$ on NCBI 2026 All closed-set. The corresponding balanced accuracies were $0.877$ and $0.890$, indicating that the model retained strong performance across families rather than only on highly represented classes. These results support FPCN's ability to classify newly released genomes from known viral families. 

\begin{table}[!t]
\centering
\caption{Evaluation of FPCN on the NCBI 2026 datasets in terms of accuracy, balanced accuracy, F1 score, recall, and precision.}
\label{tab:fpcn_results_2026}
\setlength{\tabcolsep}{7pt}
\renewcommand{\arraystretch}{1.15}

\resizebox{\columnwidth}{!}{
\begin{tabular}{lccccc}
\toprule
\textbf{Dataset} 
& \textbf{Accuracy} 
& \textbf{BA} 
& \textbf{F1} 
& \textbf{Recall} 
& \textbf{Precision} \\
\midrule

NCBI 2026 CV & 0.897 & 0.825 & 0.829 &0.825 & 0.857 \\
NCBI 2026 All CV & 0.898 & 0.823 & 0.829 & 0.823 & 0.859 \\
NCBI 2026 closed-set & 0.911 & 0.877 & 0.870 & 0.877 & 0.906\\ 
NCBI 2026 All closed-set & 0.913 & 0.890 & 0.882 & 0.880 & 0.915 \\

\bottomrule
\end{tabular}
}

\end{table}

\subsection{Ablation study on multiscale fragments}
In our proposed approach, a genome sequence is divided into different fragments using different scales. To assess the contribution of multiscale fragment integration, we retained the FPCN architecture and replaced its multiscale fragment representation with one fragment scale at a time. Each genome sequence was therefore partitioned into $15$, $17$, $21$ or $23$ fragments, with all other model components kept unchanged. The resulting models were named as PCN-$15$,  PCN-$17$, PCN-$21$ and PCN-$23$. All models were evaluated on the NCBI 2026 All temporal closet-set. FPCN achieved the highest accuracy of $0.913$ and balanced accuracy of $0.890$, exceeding the best single-scale model, PCN-17, by $0.5\%$ and $0.4\%$ respectively as presented in Table~\ref{tab:fragment_count_ablation}. At the classification level, FPCN corrected more misclassifications from the single-scale models than it introduced across all four fragment scales (see Fig.~S4). These results suggest that different fragment scales capture complementary information and that their integration yields a robust genome representation for viral family classification.

\begin{table}[!t] 
\centering
\caption{Performance comparison of single-scale and multiscale models on the NCBI 2026 All temporal closed-set evaluation.}
\label{tab:fragment_count_ablation}
\begin{tabular}{lccccc}
\toprule
\textbf{Model} &
\textbf{Accuracy} &
\textbf{BA} &
\textbf{F1} &
\textbf{Recall} &
\textbf{Precision} \\
\midrule
PCN-15 & 0.907 & 0.866 & 0.852 & 0.866 & 0.881 \\
PCN-17 & 0.908 & 0.886 & \textbf{0.886} & 0.886 & 0.919 \\
PCN-21 & 0.902 & 0.877 & \textbf{0.886} & \textbf{0.887} & \textbf{0.929} \\
PCN-23 & 0.893 & 0.881 & 0.869 & 0.881 & 0.884 \\
FPCN &\textbf{ 0.913 }& \textbf{0.890} & 0.882 & 0.880 & 0.915 \\
\bottomrule
\end{tabular}
\end{table}

\subsection{Family Error Analysis on Temporal Closed-set evaluation}
\begin{figure}[!t]
    \centering
    \includegraphics[width=0.8\linewidth]{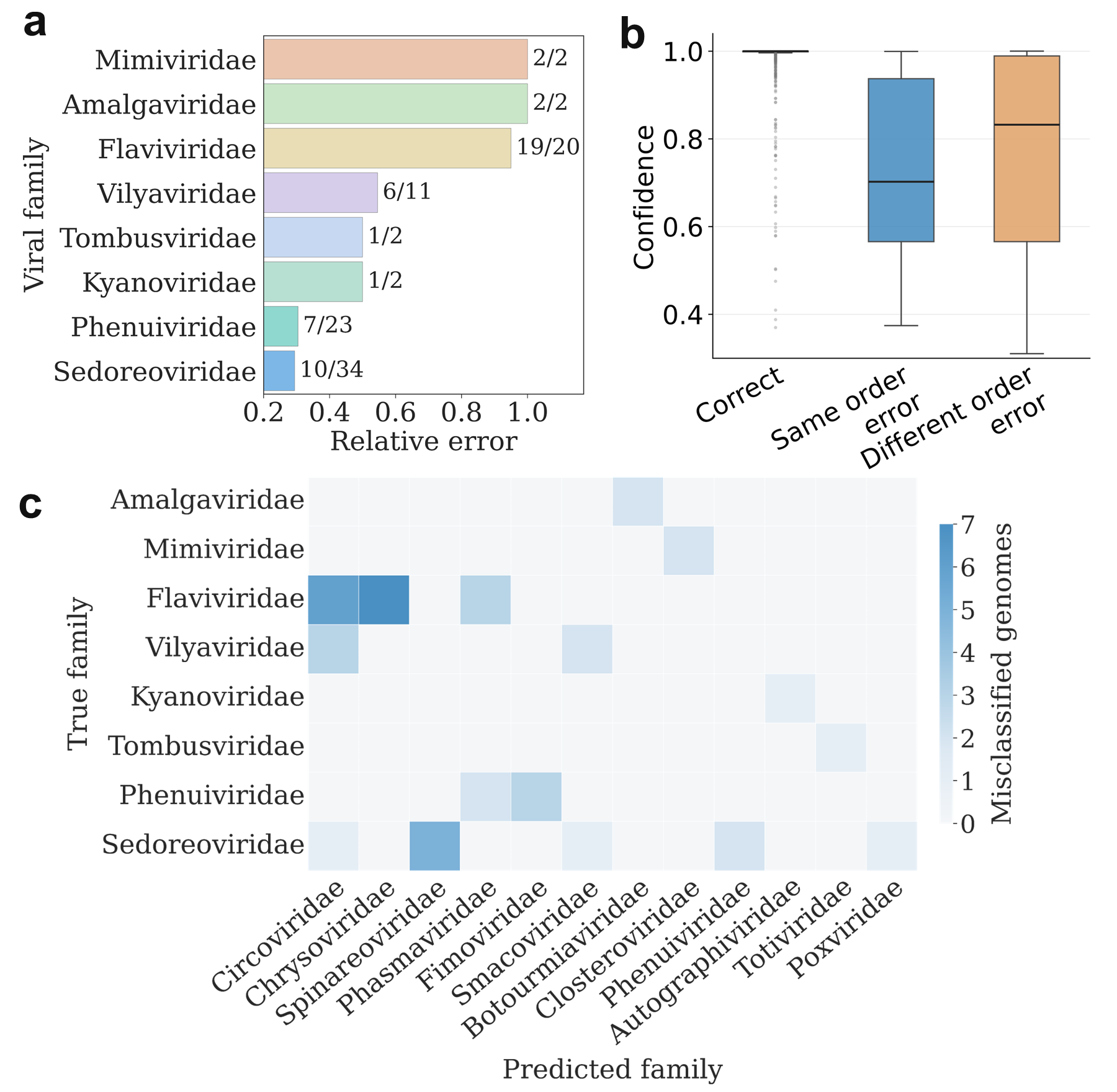}
 \caption{Family error analysis in the temporal closet-set evaluation.
    \textbf{a}, Relative errors for eight selected viral families, with labels indicating incorrectly classified genomes over total evaluated genomes. \textbf{b}, Confidence distributions of FPCN for correct family predictions, incorrect family predictions within the same order, and incorrect predictions across orders. Boxes show interquartile ranges, centre lines show medians, whiskers extend to the most extreme observations within $1.5$ interquartile ranges, and dots mark outliers.\textbf{c}, Misclassification destinations for these families, with colour intensity indicating genome counts.}
    \label{fig:external_error_analysis}
\end{figure}

To identify viral families that remained challenging for FPCN to classify, we examined misclassification errors in the temporal closet-set. Relative error was defined as the fraction of test genomes within a family that were assigned to another family. High relative errors occurred in families represented by only two external genomes, such as Mimiviridae and Amalgaviridae, as shown in Fig.~\ref{fig:external_error_analysis}a. Among families with slightly larger test sets, Flaviviridae had the highest relative error where $19$ out of $20$ genomes misclassified. Flaviviridae predictions were concentrated in Chrysoviridae and Circoviridae (Fig.~\ref{fig:external_error_analysis}c). Sedoreoviridae and Vilyaviridae had relative errors of $29.4\%$ and $54.5\%$, respectively, where most family errors were associated with assignments within the same order. Spinareoviridae accounted for half of the Sedoreoviridae errors ($5$ out of $10$). Next, we evaluated prediction confidence across classification outcomes (Fig.~\ref{fig:external_error_analysis}b). Correct predictions had a median confidence of approximately $1.000$, compared with $0.702$ for errors within the same order and $0.832$ for errors across orders. Both error groups included highly confident predictions. Confidence alone therefore did not reliably distinguish correct predictions from errors. 

\begin{figure}[!t]
    \centering
    \includegraphics[width=1\linewidth]{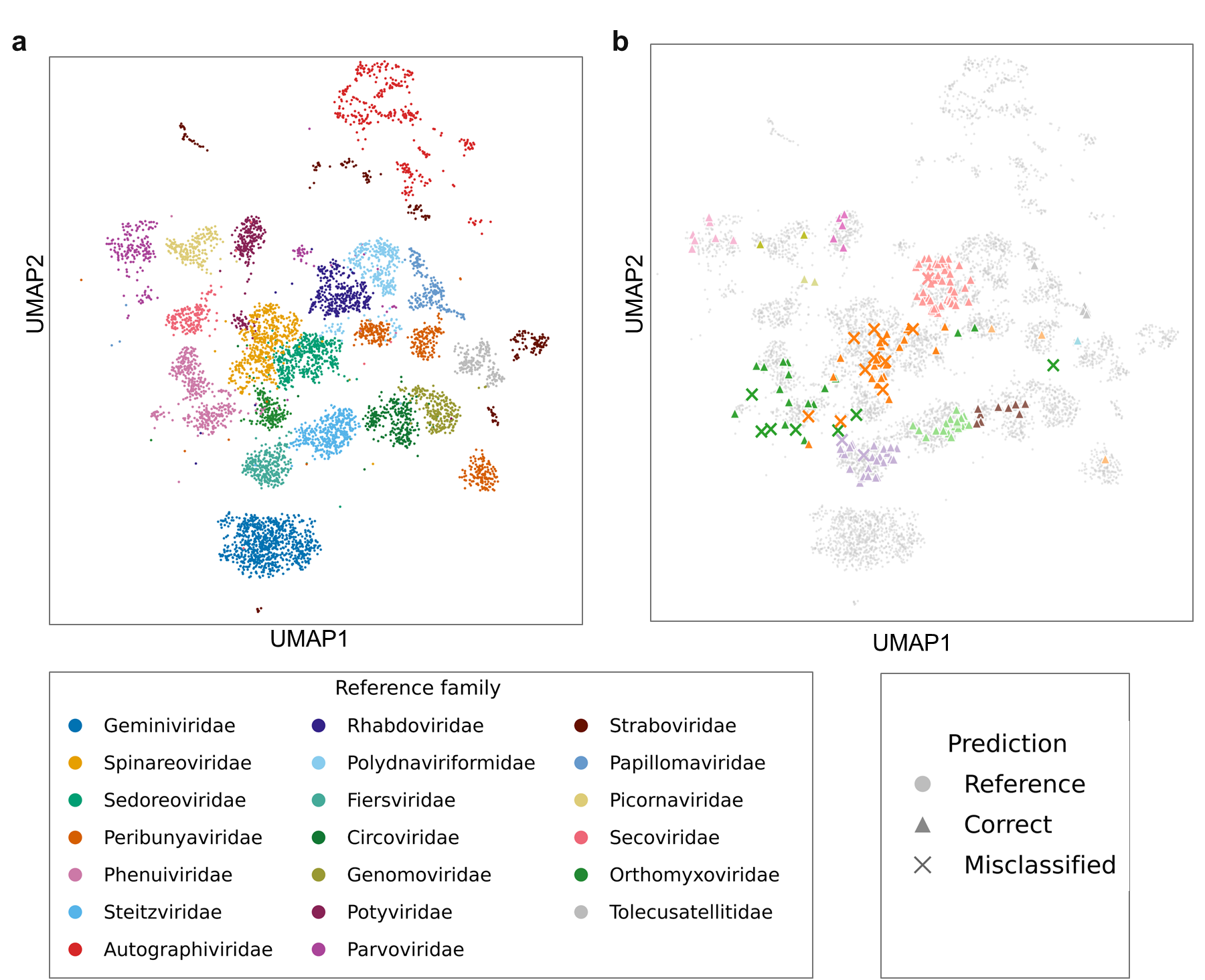}
 \caption{UMAP visualization of FPCN on temporal closet-set evaluation for $20$ viral families; \textbf{a} reference genome embeddings and \textbf{b}  temporal closet-set predictions.}
    \label{fig:umap_external_error_analysis}
\end{figure}

To explore how classification errors relate to the learned representations, we visualized FPCN genome embeddings using uniform manifold approximation and projection (UMAP) \cite{mcinnes2018umap}. For clarity, we displayed only the $20$ largest viral families as shown in Fig.~\ref{fig:umap_external_error_analysis}a,b. Families such as Geminiviridae, Fiersviridae and Steitzviridae formed distinct reference groups (Fig.~\ref{fig:umap_external_error_analysis}a). Correctly classified genomes of Fiersviridae and Steitzviridae largely coincided with their corresponding reference groups (Fig.~\ref{fig:umap_external_error_analysis}b). In contrast, Sedoreoviridae and Spinareoviridae partially overlapped, with misclassified Sedoreoviridae genomes occurring within and around the same region. This pattern is consistent with the observed confusion between these two families due to their close evolutionary relationship within Reovirales. Several misclassified Phenuiviridae genomes lay outside dense reference groups of their own family, showing that errors also occurred beyond regions of overlap between families.

\subsection{Family Error Analysis on Five-Fold Cross-Validation}
\begin{figure}[!t]
    \centering
    \includegraphics[width=0.8\linewidth]{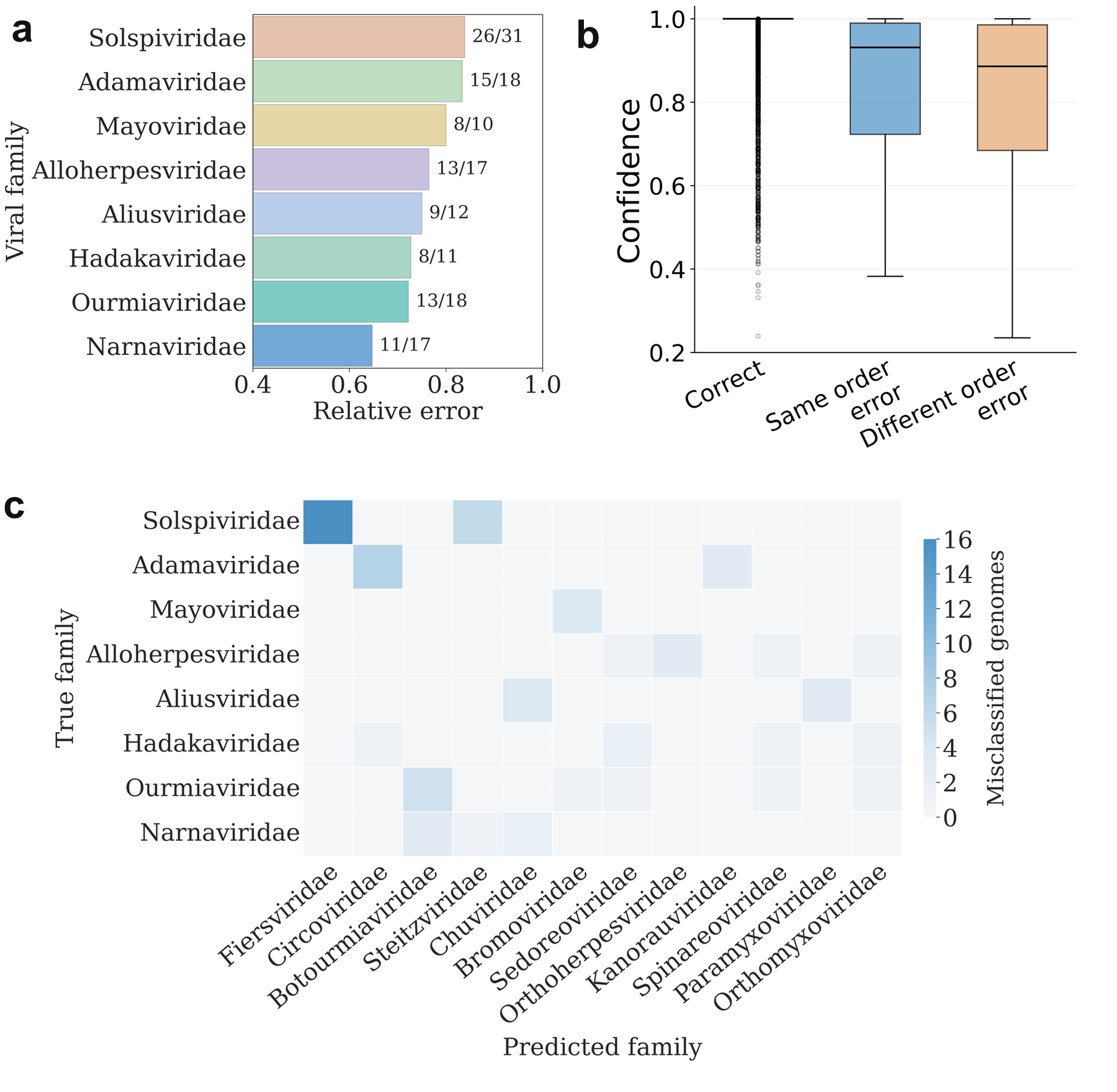}
    \caption{Family errors in five-fold cross-validation.\textbf{a,c}, Relative errors and misclassification destinations for eight selected families.
\textbf{b}, Prediction confidence by classification outcome.}
    \label{fig:internal_cv_error_analysis}
\end{figure}

For the five-fold cross-validation experiment, the highest relative errors occurred in Solspiviridae ($83.9\%$), Adamaviridae ($83.3\%$), and Mayoviridae ($80.0\%$)  as illustrated in Fig.~\ref{fig:internal_cv_error_analysis}a. We see that most misclassified Solspiviridae genomes were assigned to Fiersviridae, which belongs to the same order, Norzivirales. Adamaviridae genomes were often assigned to Circoviridae. Seven of the eight misclassified Mayoviridae genomes were assigned to Bromoviridae or Virgaviridae, which share the order Martellivirales with Mayoviridae as shown in Fig.~\ref{fig:internal_cv_error_analysis}c. Errors within the same order had a higher median confidence than errors from different orders, but the confidence distributions overlapped substantially (Fig.~\ref{fig:internal_cv_error_analysis}b). 

\begin{figure}[!t]
    \centering
    \includegraphics[width=1\linewidth]{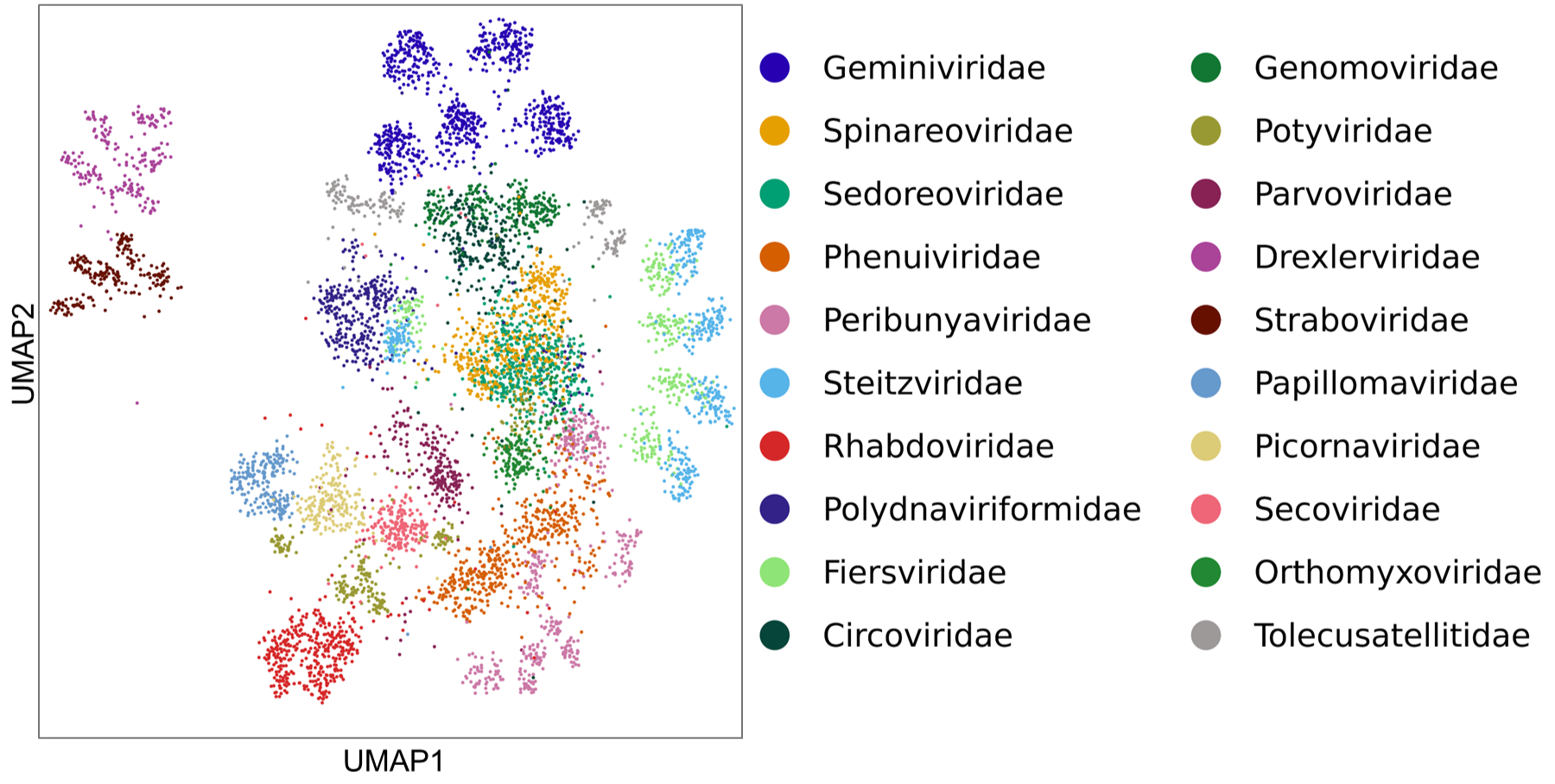}
   \caption{UMAP visualization of FPCN genome embeddings for the 20 largest viral families in internal cross-validation, coloured by family.}
    \label{fig:umap_internal_cv_error_analysis}
\end{figure}

The UMAP projection showed different patterns among the $20$ displayed families (Fig.~\ref{fig:umap_internal_cv_error_analysis}). Geminiviridae formed several separated groups, while Rhabdoviridae formed a relatively compact group.  Genomes from the same family therefore did not always cluster in a single region of the projection.

\section{Discussion} \label{sec:discussion}
The classification of viral families relies on sequence signatures found within short genomic regions or distributed across larger portions of the genome. By integrating multiscale fragments while preserving sequence order, FPCN captures complementary genomic information. Our model demonstrated higher accuracy than the reported comparison methods on the NCBI benchmarks. The performance gain is attributed to FPCN's ability to encode higher-order patterns formed across adjacent genomic regions. For the multiscale aspect of FPCN, representing each genome at different partition scales enables the model to capture sequence variation at different genome resolutions. Our ablation study demonstrates that these scales provide complementary information. Combining all four scales achieved the highest accuracy and balanced accuracy. 

The temporal evaluation extends the assessment beyond random partitions of a single dataset by testing genomes absent from the corresponding 2024 collection. FPCN maintained strong performance on these newly released genomes, indicating that the learned family-specific sequence signatures generalize beyond the genomes available during training. This is particularly important for viral classification, where newly deposited sequences may contain greater sequence variation than earlier representatives of the same family. The results suggest that FPCN captures genomic characteristics that are conserved within viral families and can generalize to later genome releases of known families. Five-fold cross-validation on NCBI 2026 datasets further showed consistent performance when trained and tested across different subsets of the newly collected genomes. 

Despite the strong predictive accuracy of FPCN, some viral families remained challenging to classify. In some difficult families, our model showed high relative errors. Particularly for families represented by few genomes and families with closely related taxonomic groups. Flaviviridae and Solspiviridae were among the most challenging cases in the temporal and cross-validation evaluations, respectively. Further analysis showed that several misclassifications occurred within the same viral order. This suggests that shared sequence characteristics among related families can make family-level classification more difficult. The overlap observed in the embedding space further supports this interpretation, as some misclassified genomes occupied regions close to those of related families.

Several directions could further improve FPCN. The relative errors observed for families with few genomes and closely related viral families suggest that taxonomic information could help distinguish similar groups. Class-balanced training or targeted augmentation may improve performance for underrepresented families. Extending FPCN to recognize previously unseen viral families is another important direction. Protein domain and genome organization features could also provide complementary biological information. Evaluation in larger and continuously updated viral datasets will be important for testing robustness as taxonomic diversity increases. Future evaluation on larger and continuously updated viral genome collections will be important for assessing the robustness of FPCN under increasingly diverse taxonomic settings.

\section{Methods}\label{sec:methods}
\subsection{Path Complex}
Path complex provides a mathematical framework for representing ordered relationships among elements of a finite set \cite{grigor2020path}. 
It underpins popular persistent path homology \cite{chowdhury2017persistent} and persistent path Laplacian  \cite{wang2023persistent}. 
Let $X$ be a finite, nonempty vertex set. An elementary $k$-path on $X$ is an ordered tuple containing $k+1$ vertices,
\begin{equation}
p^{(k)}=(x_0,x_1,\ldots,x_k),
\qquad x_j\in X.
\end{equation}
Vertices may appear more than once. A path is said to be regular if adjacent entries are distinct, meaning $x_{j-1}\neq x_j$ for every $j=1,\ldots,k$.

Suppose $\mathbb{K}$ is a coefficient field. Let $\Lambda_k(X)$ denote the vector space whose basis consists of all elementary $k$-paths on $X$. Elements of $\Lambda_k(X)$ are formal linear combinations of paths. Denote the basis element corresponding to $(x_0,\ldots,x_k)$ by $[x_0,\ldots,x_k]$. For $k\geq 0$, the boundary operator $\partial_k:\Lambda_k(X)\rightarrow\Lambda_{k-1}(X)$ is defined by
\begin{equation}
\partial_k[x_0,\ldots,x_k]
=
\sum_{j=0}^{k}(-1)^j
[x_0,\ldots,\widehat{x_j},\ldots,x_k],
\end{equation}
where $\widehat{x_j}$ denotes the omitted vertex. We use the nonaugmented convention, with $\Lambda_{-1}(X)={0}$ and $\partial_0=0$. The boundary operators satisfy $\partial_{k-1}\circ\partial_k=0$ and the spaces $\{\Lambda_k(X)\}_{k\geq 0}$, together with these maps form a chain complex:
\begin{equation}
\cdots
\xrightarrow{\partial_{k+1}}
\Lambda_k(X)
\xrightarrow{\partial_k}
\Lambda_{k-1}(X)
\xrightarrow{\partial_{k-1}}
\cdots
\xrightarrow{\partial_1}
\Lambda_0(X)
\longrightarrow 0.
\end{equation}

A path complex $\mathcal{P}$ on $X$ is a nonempty family of elementary paths satisfying the following requirement: for every included path of order $k\geq 1$, both sequences obtained by removing its first or last vertex also belong to $\mathcal{P}$ \cite{grigor2020path}. These paths are called allowed paths. Let $\mathcal{P}_{k}$ denote the set of allowed $k$-paths.
Thus, for every $k\geq 1$,
\begin{equation}
(x_0,\ldots,x_k)\in\mathcal{P}k
\quad\text{then} \quad
(x_1,\ldots,x_k)\in\mathcal{P}{k-1}
\quad\text{and}\quad
(x_0,\ldots,x{k-1})\in\mathcal{P}_{k-1}.
\end{equation}

Note that $\mathcal{P}_{-1}=\{\varnothing\}$ contains exactly one element, the empty path.

Path complexes include constructions arising from directed graphs and simplicial complexes. For a directed graph, allowed paths follow its directed edges. For a simplicial complex, fixing a total order on the vertices represents each simplex as an increasing vertex sequence. 

A path complex also gives rise to homology groups that describe its topological structure. The chain complex formed by $\Lambda_k(X)$ contains all elementary paths on $X$, so it does not distinguish the allowed paths of a particular path complex. To incorporate this information, define  
\begin{equation}
A_k(\mathcal{P})
=
\operatorname{span}_{\mathbb{K}}
\left\{
[x_0,\ldots,x_k]:
(x_0,\ldots,x_k)\in\mathcal{P}_k
\right\}.
\end{equation}
The space $A_k(\mathcal{P})$ contains the formal linear combinations of allowed paths. However, taking the boundary of an allowed path may introduce a nonallowed path through deletion of an interior vertex. Therefore the boundary operator $\partial_k$ does not necessarily map $A_{k}(\mathcal{P})$ into $A_{k-1}(\mathcal{P})$ since $\partial_{k}(A_{k}(\mathcal{P})) \not\subset A_{k-1}(\mathcal{P})$. To address this, consider the subspaces 
\begin{equation}
\Omega_k(\mathcal{P})
=
\left\{
c\in A_k(\mathcal{P}):
\partial_k c\in A_{k-1}(\mathcal{P})
\right\},
\qquad k\geq 1,
\end{equation}
with $\Omega_0(\mathcal{P})=A_0(\mathcal{P})$ and $\partial_0=0$. Since $\partial_{k-1}\circ\partial_k=0$, then boundary of every chain in $\Omega_k(\mathcal{P})$ belongs to $\Omega_{k-1}(\mathcal{P})$, i.e $\Omega_k(\mathcal{P})\subset \Omega_{k-1}(\mathcal{P})$. Consequently, we obtain a chain complex

\begin{equation}\label{eq:chain complex Q}
\cdots
\xrightarrow{\partial_{k+1}}
\Omega_k(\mathcal{P})
\xrightarrow{\partial_k}
\Omega_{k-1}(\mathcal{P})
\xrightarrow{\partial_{k-1}}
\cdots
\xrightarrow{\partial_1}
\Omega_0(\mathcal{P})
\longrightarrow 0.
\end{equation}
The $k$-th homology of (\ref{eq:chain complex Q}) is known as the $k$-th homology of $\mathcal{P}$ denoted as $H_{k}(\mathcal{P};\mathbb{K})$. In particular, for the case of a simplicial complex, path homology coincides with simplicial homology. The $k$-th Betti number is defined by
\begin{equation}
\beta_k
=
\operatorname{rank} H_k(\mathcal{P};\mathbb{K}),
\end{equation}
and therefore agrees with the corresponding simplicial Betti number.

\subsection{Path Complex Construction from Viral Genome Sequences}


Let $S$ denote a viral genome sequence of length $L$. To construct path complexes, we begin by representing each viral sequence as a multiscale fragment sequence, each sequence fragmented using the set of scales $C=\{15,17,21,23\}$. Here $c\in C$ specifies the number of fragments used to partition the genome sequence. At scale $c\in C$, we partition $S$ into $c$ consecutive non overlapping equal coordinate segments $S=\big(S^{(c)}_1,S^{(c)}_2,\ldots,S^{(c)}_c\big)$. This construction ensures that each genome sequence produces four ordered segment sets. The different fragment counts provide complementary representations of the same genome.

For each selected fragment $S_{j}^{c} ,\quad j\in \{1,2,\cdots, c\}$ with at most $512$ bases, a pretrained DNABERT-2 \cite{zhou2024dnabert} is used to encode the sequence into $d$-dimensional vector. Fragments containing at most $512$ bases are encoded directly. If a fragment exceeds the DNABERT-2 input limit, it is divided into consecutive subwindows. Each subwindow contains at most $512$ bases except possibly the final subwindow. A subwindow is valid when it contains at least one canonical token representation derived entirely from A, C, T or G positions. We average valid subwindow embeddings to form one embedding for the original segment. DNABERT-2 is then used to encode each subwindow as a $d$-dimensional vector. Segments without valid canonical tokens receive a zero vector. Within each scale, the segment embeddings define the $0$-path features. The $1$-path features are constructed from the mean embeddings of adjacent segment pairs, while $2$-path features are the average of consecutive segment triples. Each fragment scale $c\in C$ produces a scale-specific ordered path complex. The features obtained serve as input for the path complex neural network. More details can be found in the Supporting Information Section 2.

\subsection{PCNN Architecture}
The PCNN that we used learns hierarchical genome representations from each scale-specific path complex. The features from $0$-path, $1$-path, and $2$-path are first mapped to a common hidden dimension using the input projections from each rank. Stacked PCNN layers are then used to update these rank-specific representations through bidirectional information flow across adjacent path ranks. Downward messages are computed by an attention mechanism, using higher-rank path features as edge attributes, whereas upward messages are obtained by projecting the concatenated endpoint features from the lower rank. Each update is followed by residual addition and a layer normalization operation. After the final layer, a mean pooling operation is used to summarize the updated features within each path rank. The pooled $0$-, $1$- and $2$-path summaries are concatenated and transformed by a multilayer perceptron into one genome embedding for a specific scale. The same PCNN is applied to all four scales. The resulting scale-specific genome embeddings are then combined using learned softmax weights, and a linear classifier predicts the viral family.

\section{Conclusion}\label{subsec:conclusion}
This work presents Fragment Path Complex Networks (FPCN), a novel framework for viral genome classification that combines multiscale genome fragmentation, DNABERT-2 sequence embeddings, and path complex neural networks. By preserving both sequential context and higher-order relationships among genomic fragments, FPCN captures rich structural and biological information from viral genomes. Experimental results on benchmark NCBI datasets demonstrate that FPCN achieves superior classification performance compared with existing state-of-the-art methods. Furthermore, evaluations on newly collected viral genomes show strong generalization capability, indicating robustness to previously unseen sequences from known viral families. Ablation analyses further confirm that integrating information across multiple fragment scales provides more discriminative representations than single-scale approaches. These findings highlight the effectiveness of topological deep learning (TDL) for modeling genomic sequence organization. Future research will extend FPCN to open-set viral classification, explore additional biological features, and evaluate performance on larger and more diverse viral repositories. Overall, FPCN offers a scalable and accurate approach for next-generation viral genome analysis and taxonomy.
The proposed approach can be applied to gene replacement, protein-nucleic acid binding analysis, and other tasks involving DNA/RNA.

\section*{Supporting Information}
Dataset descriptions, sample counts, and genome length distributions are provided in the Supporting Information. Details of multiscale path complex construction, feature encoding, and model implementation are included. Evaluation metrics and benchmark performance comparisons are also presented. Additional analyses include UMAP visualizations of genome embeddings and prediction changes between FPCN and single-scale models.

\section*{Conflicts of Interest}
The authors declare no competing financial interests

\section*{Data availability}

The original viral genome records analysed in this study were obtained from the NCBI Virus database (\url{https://www.ncbi.nlm.nih.gov/labs/virus/vssi/}). The NCBI 2026 datasets were accessed on 10 August 2026. The preprocessed NCBI benchmark datasets are available through Zenodo at \url{https://zenodo.org/records/19655917}. The processed NCBI 2026 datasets generated for this study are available through Zenodo at \url{https://zenodo.org/records/22849345}.

\section*{Code availability}

The source code used to implement FPCN is available at \url{https://github.com/wachiraa26/FPCN}.

\section*{Acknowledgments}
 This work was supported in part by NIH Grant R01AI164266 and the Georgia Research Alliance.  
%
\bibliographystyle{unsrt}
\bibliography{references}

\end{document}